\documentclass[12pt]{article}
\begin{document}


\title{Supersymmetric dark energy}
\author{
Neven Bili\'c\thanks{Electronic mail:
bilic@thphys.irb.hr}
 \\
Rudjer Bo\v skovi\'c Institute,
POB 180, HR-10002 Zagreb, Croatia 
}

\maketitle

\begin{abstract}
We study a noninteracting supersymmetric model in de Sitter spacetime.
A soft supersymmetry breaking induces a nonzero vacuum energy density.
 A  short distance cut-off of the order of Planck length
provides a matching between the vacuum energy density and the cosmological constant
related to the de Sitter expansion parameter.

\end{abstract}



\section{Introduction}

It is generally accepted that the cosmological constant term
which was introduced ad-hoc in the Einstein-Hilbert action
is actually related to the vacuum energy density of matter fields.
The vacuum energy density estimated in a simple quantum field theory
is by about 120 orders of magnitude larger than the value required by
astrophysical and cosmological observations. 
On the other hand, in a field theory with exact supersymmetry the
vacuum energy, and hence the cosmological constant (CC), is equal to zero as the
contributions  of fermions and bosons to the vacuum energy precisely
cancel \cite{weinberg}. 

A nonzero CC implies the de Sitter 
symmetry group of spacetime rather than the Poincar\'e group which is the spacetime symmetry group of
an exact supersymmetry. Based on observational evidence for an
  accelerating expansion \cite{perlmutter,bennett,spergel},
 the vacuum energy density
dominates
the total energy density today.
Hence, the spacetime today is close to de Sitter approaching asymptotically
a de Sitter universe 
with metric
\begin{equation}
ds^2= dt^2 - a(t)^2 d\vec{x}\,^2,
\label{eq000}
\end{equation}
where $a=e^{Ht}$ and
$H$ is a constant.
This metric describes empty space 
with cosmological constant $\Lambda=8\pi G \rho_\Lambda$,
where the  vacuum energy density $\rho_\Lambda$ is related to $H$ by the Friedman equation:
\begin{equation}
H^2=\left(\frac{\dot{a}}{a}\right)^2=\frac{8\pi G}{3}\rho_\Lambda .
\label{eq213}
\end{equation}

The structure of de Sitter spacetime
automatically breaks the supersymmetry \cite{banks}.
Conversely, a low energy supersymmetry breaking
could in principle generate a  nonzero CC of an
acceptable magnitude.
 Unfortunately, the scale of supersymmetry breaking required by the particle physics phenomenology
must be of the order of 1 TeV or larger implying a CC too large
by about 60 orders of magnitude.
However, 
some nonsupersymmetric models with equal number of boson and fermion degrees of
 freedom have been constructed \cite{kamenshchik} so that
 all the divergent contributions to the vacuum energy density cancel and a small finite contribution
can be made comparable with the observed value
of the CC.

In this paper we investigate the fate of vacuum energy
when an unbroken supersymmetric model 
is embedded in asymptotically de Sitter spacetime.
We do not claim that our model describes a realistic scenario
but it is tempting to speculate along the the following lines.
Large scale observations reveal that dark and baryonic matter gravitationally cluster occupying
relatively small volume 
of space in the form of poor and reach clusters connected by filaments and sheets \cite{bond}.
Most of the volume is occupied by large scale voids almost empty of both baryon and dark matter.

Our working assumption is that voids contain no matter apart from fluctuations of
a supersymmetric vacuum as a relict of symmetry breaking in the early universe.
The early universe with exact supersymmetry
underwent a set of symmetry breaking phase transitions. 
Before the supersymmetry breaking, domains with different vacua may have been formed 
with domain walls separating the domains \cite{morris}. Then, at a later time,
supersymmetry breaking took place in some if the domains, which thereafter remained  populated by dark and baryonic
matter. Our main assumption is  that supersymmetry remains unbroken in the voids.
However, since the global geometry is de Sitter,
the lack of Poincare symmetry will lift the Fermi-Bose degeneracy
and the energy density of vacuum fluctuations will
be nonzero. This type of ``soft'' supersymmetry breaking is similar to the supersymmetry breaking at finite temperature 
where the Fermi-Bose degeneracy is lifted by quantum statistics \cite{das,girardello}.

The remainder of the paper is organized as follows. In section \ref{model} we describe our model.
The calculations and results are presented in section \ref{calculations} and concluding remarks
in section \ref{conclude}.

\section{The model}
\label{model}
Here we consider a noninteracting Wess-Zumino supersymmetric model with $N$ species and calculate the energy density of vacuum fluctuations
in de Sitter spacetime.
 In general, the supersymmetric Lagrangian $\cal L$ 
for $N$ chiral superfields has the form \cite{bailin}
\begin{equation}
{\cal L} =  \sum_i \Phi_i^\dag \Phi_i |_D +W(\Phi)|_F + \rm h.c.\, ,
\label{eq001}
\end{equation}
where the index $i$ distinguishes the various left chiral superfields $\Phi_i$
and $W(\Phi)$ denotes the superpotential for which we take
\begin{equation}
W(\Phi) = \frac{1}{2}  \sum_i m_i \Phi_i\Phi_i \, .
\label{eq002}
\end{equation}
Eliminating auxiliary fields by equations of motion
the Lagrangian (\ref{eq001}) may be recast in the form
\begin{equation}
{\cal L} =  \partial_\mu \phi_i^\dag \partial^\mu\phi_i -m_i^2 |\phi_i|^2
+\frac{i}{2}\bar{\Psi}_i\gamma^\mu \partial_\mu\Psi_i -\frac{1}{2} m_i\bar{\Psi}_i\Psi_i \, ,
\label{eq003}
\end{equation}
where $\phi_i$ are the complex scalar and $\Psi_i$ the Majorana spinor fields.
For simplicity, from now on we suppress the dependence on the species index $i$.

Next we assume a curved background spacetime geometry with metric $g_{\mu\nu}$.
Spinors in curved spacetime are conveniently treated using the
so called vierbein formalism.
The metric is decomposed as
\begin{equation}
g_{\mu\nu}(x) = {e^a}_\mu {e^b}_\nu \eta_{ab},
\label{eq101}
\end{equation}
where the set of coefficients ${e^a}_\mu$ is  called the {\em vierbein}.
The action  
 may be written as 
\begin{equation}
S= \int d^4x \sqrt{-g}  ( {\cal L}_B + {\cal L}_F) ,
\label{eq004}
\end{equation}
where ${\cal L}_B$ and ${\cal L}_F$ are the boson and fermion Lagrangians, respectively.
The Lagrangian for a complex scalar field may be expressed as the sum of the
Lagrangians for two real fields 
\begin{equation}
{\cal L}_B =  \frac{1}{2}\sum_{i=1}^2 \left(g^{\mu\nu}\varphi^i_{,\mu} \varphi^i_{,\nu} 
-m^2 \varphi^{i\,2}\right).
\label{eq005}
\end{equation}
The fermion part is given by \cite{birrell}
\begin{equation}
{\cal L}_F = 
\frac{i}{4}\left(\bar{\Psi}\tilde{\gamma}^\mu\Psi_{;\mu}-
\bar{\Psi}_{;\mu}\tilde{\gamma}^\mu \Psi\right)-
\frac{1}{2} m\bar{\Psi}\Psi ,
\label{eq006}
\end{equation}
where $\tilde{\gamma}^\mu$ are the curved spacetime gamma matrices 
\begin{equation}
\tilde{\gamma}^\mu = e^\mu_a \gamma^a ,
\label{eq007}
\end{equation}
with  ordinary Dirac gamma matrices denoted by $\gamma^a$, and ${e_a}^\mu$ is the inverse of the vierbein.
The covariant derivatives of the spinor are defined as
\begin{equation}
\Psi_{;\mu}=\Psi_{,\mu} - \Gamma_\mu \Psi ,
\label{eq008}
\end{equation}
\begin{equation}
\bar{\Psi}_{;\mu}=\bar{\Psi}_{,\mu} +\bar{\Psi} \Gamma_\mu \, ,
\label{eq009}
\end{equation}
where 
\begin{equation}
\Gamma_\mu= \frac{1}{8}{\omega_\mu}^{ab}[\gamma^a,\gamma^b] \, ,
\label{eq010}
\end{equation}
with the spin connection \cite{parker}
\begin{equation}
{\omega_\mu}^{ab}=-\eta^{bc}{e_c}^\nu ({e^a}_{\nu,\mu}-\Gamma^\lambda_{\mu\nu}{e^a}_\lambda) .
\label{eq011}
\end{equation}
In FRW metric the vierbein is diagonal and in spatially flat FRW spacetime takes a simple form
\begin{equation}
 {e^a}_\mu = {\rm diag} (1,a,a,a) 
\label{eq111}
\end{equation}
where $a=a(t)$ is the cosmological expansion scale.

\section{Calculation of the vacuum energy density}
\label{calculations}

It is convenient to work in the conformal frame with metric
\begin{equation}
ds^2= a(\eta)^2(d\eta^2 - d\vec{x}\,^2).
\label{eq012}
\end{equation}
where the  proper time $t$ of the isotropic observers is related to the conformal time $\eta$ as
\begin{equation}
dt= a(\eta) d\eta.
\label{eq013}
\end{equation}
In particular, we will be interested in de Sitter spacetime with
\begin{equation}
a=e^{Ht} = -\frac{1}{H\eta} .
\label{eq113}
\end{equation}

In order to calculate the energy density of the vacuum fluctuations 
we need the vacuum expectation value of the Hamiltonian.
The Hamiltonian may be expressed as the sum of the boson and fermion parts
\begin{equation}
{\cal H} = {\cal H}_B +{\cal H}_F\, .
\label{eq025}
\end{equation}
From (\ref{eq004}-\ref{eq006}) with  metric  (\ref{eq012}) we obtain
\begin{equation}
{\cal H}_B =  \sum_{i=1}^2\left(\frac{1}{2a^2}(\partial_\eta\varphi^i)^2+ 
\frac{1}{2a^2}(\nabla \varphi^i)^2 +m^2 \varphi^{i\,2}\right),
\label{eq026}
\end{equation}
\begin{equation}
{\cal H}_F = 
-i\frac{1}{4a^4}\left(\bar{\psi}\gamma^j\partial_j\psi-
(\partial_j\bar{\psi})\gamma^j \psi\right)
+\frac{1}{2a^3} m\bar{\psi}\psi .
\label{eq027}
\end{equation}


Consider first the contribution of the scalar fields.
Each real  scalar field operator is decomposed as
\begin{equation}
\varphi(\eta , \vec{x}) = \sum_{\vec{k}} a^{-1}\left( \chi_k(\eta)e^{i \vec{k}\vec{x}} a_k
+\chi_k(\eta)^*e^{-i \vec{k}\vec{x}} a_k^\dag \right) ,
\label{eq014}
\end{equation}
where $a_k$ and $a_k^\dag$ are the annihilation and creation operators, respectively. 
The function $\chi_k$ satisfies the field equation
\begin{equation}
{\chi}^{\prime\prime}_k+ (m^2a^2+k^2- a''/a) \chi_k=0,
\label{eq015}
\end{equation}
where $'$ denotes a derivative with respect to the conformal time $\eta$.
In massless case the exact solutions to this equation may easily be found \cite{birrell}.
In particular, in de Sitter spacetime $a''/a=1/\eta^2$, and one finds positive frequency solutions \cite{danielsson}
\begin{equation}
\chi_k =  \frac{1}{\sqrt{2Vk}}e^{-ik\eta}\left(1-\frac{i}{k\eta}\right).
\label{eq016}
\end{equation}
The operators $a_k$ associated to these solutions annihilate the 
adiabatic vacuum  in the asymptotic past (Bunch-Davies vacuum) \cite{birrell,jacobson}.

If $m\neq 0$  solutions to (\ref{eq015}) may be constructed by making use of the 
WKB ansatz 
\begin{equation}
\chi_k(\eta) =  \frac{1}{\sqrt{2VaW_k(\eta)}}e^{-i\int^\eta a W_k(\tau) d\tau} ,
\label{eq017}
\end{equation}
where the function $W_k$ may be found by solving (\ref{eq015}) iteratively up to an arbitrary 
order in adiabatic expansion \cite{parker}. For our purpose we need the solution up to
the  2nd order only which reads
\begin{equation}
W_k =  \omega_k +\omega^{(2)} ,
\label{eq117}
\end{equation}
where 
\begin{equation}
\omega_k =  \sqrt{m^2+k^2/a^2} .
\label{eq217}
\end{equation}
The general expression for the second order term 
is \cite{parker}
\begin{equation} 
\omega^{(2)}=
 - \frac{3}{8}\frac{1}{\omega_k}\frac{\dot{a}^2}{a^2}
- \frac{3}{4}\frac{1}{\omega_k}\frac{\ddot{a}}{a}
- \frac{3}{4}\frac{k^2}{a^2\omega_k^3}\frac{\dot{a}^2}{a^2} 
+\frac{1}{4}\frac{k^2}{a^2\omega_k^3}\frac{\ddot{a}}{a}
+\frac{5}{8}\frac{k^4}{a^4\omega_k^5}\frac{\dot{a}^2}{a^2} \, ,
\label{eq317}
\end{equation}
where overdot denotes a derivative with respect to  $t$.
For  de Sitter spacetime, using (\ref{eq113}) we obtain 
\begin{equation}
W_k =  \omega_k -\frac{H^2}{\omega_k}\left[1+{\cal{O}}(m^2/\omega_k^2)\right] ,
\label{eq417}
\end{equation}

We can calculate now the vacuum expectation value of the boson Hamiltonian.
Using the properties of $a$ and $a^\dag$ and replacing the sum over momenta
 by an integral
in the usual way
\begin{equation}
\sum_{\vec{k}} =  V\int \frac{d^3k}{(2\pi)^3} \, ,
\label{eq128}
\end{equation}
from (\ref{eq014}) we find
\begin{equation}
<{\cal H}_B> =  \frac{V}{a^4}\int \frac{d^3k}{(2\pi)^3}\left( |\chi_k^{\,\prime}|^2+ 
a^2 \omega_k^2| \chi_k|^2\right).
\label{eq028}
\end{equation}
Using (\ref{eq017})  with (\ref{eq417}) we obtain
\begin{equation}
<{\cal H}_B>=\frac{1}{a^3}\int \frac{d^3k}{(2\pi)^3}\left[\omega_k+\frac{1}{2}\frac{H^2}{\omega_k}
\left(1+ \frac{m^2}{\omega_k^2} -\frac{H^2}{\omega_k^2}+{\cal O}(\omega_k^{-4})\right) \right] .
\label{eq030}
\end{equation}
The first term in square brackets is identical to the flat spacetime result. The second term
is a quadratically divergent contribution due to de Sitter geometry, the next two terms are 
logarithmically divergent, and the rest is finite.


Next we  proceed to quantize the fermions.   The 
Dirac equation in curved spacetime may be derived from (\ref{eq006}).
 Specifically for a spatially flat FRW metric
we obtain 
 \begin{equation}
i\gamma^0\left(\partial_0+\frac{3}{2}\frac{\dot{a}}{a}\right)\Psi+
i\frac{1}{a}\gamma^j\partial_j\Psi
-m\Psi=0.
\label{eq018}
\end{equation}
It is  convenient to rescale  the
Majorana fermion field $\Psi$  as
\begin{equation}
\Psi=a^{-3/2}\psi .
\label{eq019}
\end{equation}
and introducing the conformal time
we obtain for $\psi$   the usual flat spacetime Dirac equation 
\begin{equation}
i\gamma^0\partial_\eta\psi+
i\gamma^j\partial_j\psi
-am\psi=0 ,
\label{eq020}
\end{equation}
with time dependent effective mass $am$.
The quantization of $\psi$ is now straightforward
\cite{baacke,cherkas}.
The Majorana field $\psi$ may be decomposed as usual
\begin{equation}
\psi(\eta , \vec{x}) = \sum_{\vec{k},s} \left( u_{ks}(\eta)e^{i \vec{k}\vec{x}} b_{ks}
+v_{ks}(\eta)e^{-i \vec{k}\vec{x}} b_{ks}^\dag \right),
\label{eq021}
\end{equation}
where the spinor $u_{ks}$ may be expressed as
\begin{equation}
u_{ks}= \frac{1}{\sqrt{V}}\left(
\begin{array}{c}
 (i\zeta_k^{\,\prime}+am\zeta_k ) \phi_s \\
\vec{\sigma} \vec{k} \, \zeta_k \phi_s
\end{array}
\right).
\label{eq022}
\end{equation}
Here, the two-spinors $\phi_s$ are the helicity eigenstates which may be chosen as
\begin{equation}
\phi_+ =\left( \begin{array}{c}
 1 \\
0
\end{array}
\right); \hspace{1cm} 
\phi_- =
\left(\begin{array}{c}
 0 \\
1
\end{array}
\right).
\label{eq023}
\end{equation}
The spinor $v_{ks}$ is related to $u_{ks}$ by charge conjugation 
\begin{equation}
v_{ks}= i \gamma^0\gamma^2 (\bar{u}_{ks})^T .
\label{eq122}
\end{equation}
The norm of the spinors may be easily calculated
\begin{equation}
\bar{u}_{ks}u_{ks}=-\bar{v}_{ks}v_{ks}=
\frac{1}{V}(am \zeta_k^*-i\zeta_k^{*\prime})
(am \zeta_k+i\zeta_k')-\frac{1}{V}k^2|\zeta_k|^2.
\label{eq422}
\end{equation}
The mode functions $\zeta_k$ satisfy the equation
\begin{equation}
\zeta_k''+ (m^2a^2+k^2-im a') \zeta_k=0.
\label{eq024}
\end{equation}
In addition, the functions $\zeta_k$ satisfy the condition
\cite{cherkas}
\begin{equation}
k^2|\zeta_k|^2+ (am \zeta_k^*-i\zeta_k^{*\,\prime})
(am \zeta_k+i\zeta_k^{\,\prime})=C_1^2 .
\label{eq124}
\end{equation}
It may be easily verified that the left-hand side of this equation is a constant of motion
of equation (\ref{eq024}). 
The constant $C_1$ is fixed by the normalization of the spinors 
and by the  initial conditions. A natural assumption is that
at $t=0$ ($\eta=-1/H$, $a=1$) the solution behaves as a plane wave
$\zeta_k=C_2e^{-iEt}$,
where $E=\sqrt{k^2+m^2}$.
 This gives $\zeta_k(0)=C_2$, $\zeta_k^{\,\prime}(0)=-iC_2E$,
and hence  $C_1^2=2C_2^2E (m+E)$. From (\ref{eq422}) and (\ref{eq124}) we obtain 
\begin{equation}
\bar{u}_{ks}u_{ks}=-\bar{v}_{ks}v_{ks}=\frac{1}{V}(C_1^2
-2k^2|\zeta_k|^2) ,
\label{eq222}
\end{equation}
which at $t=0$ reads  
\begin{equation}
\bar{u}_{ks}u_{ks}=-\bar{v}_{ks}v_{ks}=
C_1^2\frac{m}{VE} \, .
\label{eq322}
\end{equation}
For $ C_1^2=1$ this coincides with the standard  flat spacetime normalization \cite{birrell}.

In massless case the solutions to (\ref{eq024}) are plane waves.
For $m\neq 0$ two methods have been used to solve (\ref{eq024}) for a general spatially flat FRW spacetime:
a) expanding in negative powers of $\sqrt{m^2+k^2}$ and solving a recursive set of differential equations
\cite{baacke} b) using  a WKB ansatz similar to  (\ref{eq017}) and the adiabatic expansion \cite{cherkas}.

By making use of the decomposition (\ref{eq021}) and the standard anti-commuting
properties of the creation and annihilation operators, 
the vacuum expectation value of the fermion 
Hamiltonian (\ref{eq027}) may be written as
\begin{equation}
 <{\cal H}_F> = \frac{1}{2a^4}\sum_{\vec{k},s} \bar{v}_{ks}
 (am-\vec{k}\,\vec{\gamma})
v_{ks} \, .
\label{eq031}
\end{equation}
Evaluating the expression under the sum and replacing the sum with an integral as in (\ref{eq128})
we obtain
\begin{equation}
<{\cal H}_F>=\frac{1}{a^4}\int \frac{d^3k}{(2\pi)^3}\left[ik^2
(\zeta_k\zeta_k^{*\prime} -\zeta_k^*\zeta_k')-am \right].
\label{eq032}
\end{equation}
The expression in square brackets was calculated in \cite{baacke} for a spatially flat
FRW metric. We quote their result for the divergent contribution:
\begin{equation}
<{\cal H}_F>_{\rm div}=\frac{1}{a^4}\int \frac{d^3k}{(2\pi)^3}\left[
-E -\frac{(a^2-1)m^2}{2E}
+\frac{(a^2-1)^2m^4}{8E^3}
+\frac{(a')^2m^2}{8E^3}
 \right].
\label{eq033}
\end{equation}
Note that the first three terms in square brackets are identical to the first three
terms in the expansion of $a\omega_k=\sqrt{E^2 + a^2m^2-m^2}$ in powers of
$E^{-2}$.
Hence we can write
\begin{equation}
<{\cal H}_F>_{\rm div}=\frac{1}{a^3}\int \frac{d^3k}{(2\pi)^3}\left[
-\omega_k +\frac{(a')^2m^2}{8a^4\omega_k^3}
+{\cal O}(\omega_k^{-5})
 \right].
\label{eq034}
\end{equation}
The first term in square brackets is precisely the  flat spacetime vacuum energy
of the fermion field. 
The second term
is a logarithmically divergent contribution due to the FRW geometry and 
the last term is finite and vanishes in the flat-spacetime limit
$ a' \rightarrow 0$.
Note  that, as opposed to bosons,  there is no quadratic divergence
of the type $H^2/\omega_k$.


Assembling the boson and fermion contributions, 
the final expression for the vacuum energy density 
of each chiral supermultiplet is
\begin{equation}
\rho_{\rm vac}=<{\cal H}_F+{\cal H}_B>=\frac{1}{a^3}\int \frac{d^3k}{(2\pi)^3}\frac{1}{2}\frac{H^2}{\omega_k}\left[1
+ \frac{5}{4}\frac{m^2}{\omega_k^2} -\frac{H^2}{\omega_k^2}+{\cal O}(\omega_k^{-4}) \right].
\label{eq035}
\end{equation}
The dominant contribution comes from the first term which diverges quadratically.
To make the result finite we  change the integration variable  to the physical momentum 
$p=k/a$ and introduce a cutoff of the order of the Planck mass 
$\Lambda_{\rm cut}\sim  m_{\rm Pl}$. The leading term yields the total 
energy density of the vacuum fluctuations
\begin{equation}
\rho_{\rm vac}=\frac{NH^2}{4\pi^2}\int_0^{\Lambda_{\rm cut}}\!\!
p\,d\!p \left(1+{\cal O}(p^{-2})\right)
\cong \frac{NH^2\Lambda_{\rm cut}^2}{8\pi^2}
\left(1+{\cal O}(\Lambda_{\rm cut}^{-2}\ln{\Lambda_{\rm cut}})\right),
\label{eq036}
\end{equation}
where $N$ is the number of chiral species. 
Note that the leading term in (\ref{eq035}) is due to bosons; 
fermions only provide a precise cancellation of the
quartically divergent flat spacetime vacuum term. 

If the Einstein cosmological term is precisely zero then the only source 
of the vacuum energy density $\rho_\Lambda$ are the vacuum fluctuations of matter fields.
Hence we identify $\rho_{\rm vac}=\rho_\Lambda$ and 
if we compare (\ref{eq036}) with
the Friedman equation (\ref{eq213}) 
we find that our  cutoff should satisfy
\begin{equation}
\Lambda_{\rm cut}\cong \sqrt{\frac{3\pi}{N}}\, m_{\rm Pl} \, .
\label{eq037}
\end{equation}

It is worthwhile to note that several  approaches 
\cite{shapiro,sloth2,gurzadyan,padmanabhan,cohen,maggiore,sloth,bilic2} 
with substantially 
different underlying philosophy
lead to results similar to (\ref{eq036}).
Fluctuations
in the energy density of a scalar field were estimated \cite{padmanabhan}
yielding the same  order of magnitude as in (\ref{eq036}). There,
a cosmological horizon radius $R_H=1/H$ was employed as
 a long distance cutoff. This idea is similar in spirit with 
\cite{cohen} where an upper bound 
\begin{equation}
\rho_{\rm vac}\cong \Lambda_{\rm UV}^4\leq \frac{3}{8\pi}\frac{m_{\rm Pl}^2}{L^2} 
 \label{eq038}
\end{equation}
was proposed from  a holographic principle. 
Here, $\Lambda_{\rm UV}$ and $L$ denote the ultraviolet and long distance cutoffs, respectively.
Our result would saturate the holographic bound (\ref{eq038})
 provided  $L=1/H$.

In recent papers \cite{maggiore,sloth} a  residual quadratic contribution of the
form $H^2 \Lambda_{\rm cut}^2$ has been found after canceling the flat spacetime contribution.
In particular, the work \cite{maggiore} presents a  calculation
of the zero-point energy using only a massless boson field and obtains two types of contributions: 
the  quadratic  and the quartic type
$\Lambda_{\rm cut}^4$. 
Then, the quartic contributions to  CC was simply canceled
 by hand on the basis of the procedure used previously in
the literature with the so-called ADM mass. As we have demonstrated here  
(see also \cite{bilic2}),
in a supersymmetric world such a cancellation by fiat
is unnecessary because  the cancellation 
between bosons and fermions of all (not only quartically divergent) 
flat-spacetime contributions is naturally provided by supersymmetry.

Another important point \cite{maggiore,bilic2} is 
that the vacuum fluctuations cannot  be interpreted as a part of CC because
the vacuum fluctuations do not yield the equation of state
$p=-\rho$, as a consequence of the energy
momentum tensor not having a CC form.
 This behavior was already observed in flat space time if a three dimensional
cutoff regularization was employed \cite{ossola}.
A covariant regularization in flat space time should
yield the vacuum energy momentum tensor of the form
$T_{\mu\nu}=\rho \eta_{\mu\nu}$. Naively, in curved spacetime one would  
generalize this to the CC form $T_{\mu\nu}=\rho g_{\mu\nu}$.
However, since a curved geometry involves
the Riemann tensor and its covariant derivatives 
we may  expect the energy momentum tensor at linear curvature order to be
of the form $T_{\mu\nu}=(\alpha+\beta R ) g_{\mu\nu}+\gamma R_{\mu\nu}$ 
where $\alpha$,  $\beta$, and $\gamma$ are constants that do not depend on curvature.
In reference \cite{bilic3} the vacuum contribution to the energy momentum tensor
has been investigated using an explicitly covariant regularization scheme
for an unbroken supersymmetric model 
embedded in a general curved spacetime.
One loop contributions to the
effective potential were calculated  using a covariant UV cutoff
in an approach similar to Sobreira et al \cite{sobreira}.

\section{Conclusion}
\label{conclude}
We have found that the leading term in the energy density of vacuum fluctuations is of the order $H^2 m_{\rm Pl}^2$ if we impose a short distance cutoff of the order $m_{\rm Pl}^{-1}$.
  In this way, if we require that the de Sitter expansion parameter $H$ equals the Hubble parameter today,
 the model provides  a phenomenologically acceptable value
of the vacuum energy.
We have also found that a consistency with the Friedman equation
implies that a natural cutoff must be inversely proportional to $\sqrt{N}$.
A similar natural cutoff  has been recently proposed in order to
resolve the so called species problem of black-hole entropy \cite{dvali}.

\subsection*{Acknowledgments}
I wish to thank B.\ Guberina, A.Y.\ Kamenshchik, H.\ Nikoli\'c, J.\ Sola, and H.\ \v Stefan\v ci\'c for useful discussions
and comments.
This work was supported by the Ministry of Science,
Education and Sport
of the Republic of Croatia under contracts No. 098-0982930-2864 and 
partially supported by the ICTP-SEENET-MTP grant PRJ-09 ``Strings and Cosmology`` 
in the frame of the SEENET-MTP Network.

\end{document}